\documentclass{aa}  
\usepackage{natbib}
\bibpunct{(}{)}{;}{a}{}{,}
\usepackage{graphicx}
\usepackage{txfonts}
\begin{document}

   \title{Formation of Magnetic Switchbacks via expanding Alfv\'{e}n Waves}

   \author{T.A. Bowen
          \inst{1}
          \and
          A. Mallet
          \inst{1}
          \and 
          C. I. Dunn \inst{1\&2}  
          \and
          J. Squire\inst{3}
          \and B.D.G. Chandran\inst{4}
                    \and R. Meyrand\inst{3}
          \and N. Davis\inst{4}
           \and T. Dudok de Wit\inst{5,6}
          \and S.D. Bale\inst{1\&2}
          \and S.T. Badman\inst{7}
             \and N. Sioulas\inst{1}
         }

   \institute{Space Sciences Laboratory, University of California,7 Gauss Way, Berkeley, CA 94720\\
   \email{tbowen@berkeley.edu}
         \and
             Physics Department, University of California,
            Berkeley, CA 94720
             \and
                    Department of Physics, University of Otago, 730 Cumberland St., Dunedin 9016, New
                    Zealand
                    \and
                   Department of Physics \& Astronomy, University of New Hampshire, Durham, NH 03824, USA
                   \and
                   LPC2E, CNRS and University of Orl\'eans, Orl\'eans, France
                   \and   ISSI, Bern, Switzerland
                    \and  Harvard \& Smithsonian Center for Astrophysics, Cambridge, MA, USA}

 
  \abstract
   {Large-amplitude inversions of the solar wind's interplanetary magnetic field have long been documented; however, observations from the Parker Solar Probe (PSP) mission have renewed interest in this phenomenon as such features, often termed switchbacks, may constrain both the sources of the solar wind as well as in-situ nonlinear dynamics and turbulent heating.}
   {We aim to show that magnetic field fluctuations in the solar wind are consistent with Alfv\'{e}nic fluctuations that naturally form switchback inversions in the magnetic field through expansion effects.}
   {We examine PSP observations of the evolution of a single stream of solar wind in a radial scan from PSP's tenth perihelion encounter from $\approx 15-50 R_\odot$. We study the growth and radial scaling of normalized fluctuation amplitudes in the magnetic field, $\delta B/B$, within the framework of spherical polarization. We compare heating rates computed via outer-scale decay from consideration of wave-action to proton heating rates empirically observed through considering adiabatic expansion.}
   {We find that the magnetic field fluctuations are largely spherically polarized and that the normalized amplitudes of the magnetic field, $\delta B/B$, increases with amplitude. The growth of the magnetic field amplitude leads to switchback inversions in the magnetic field. While the amplitudes do not grow as fast as predicted by the conservation of wave action, the deviation from the expected scaling yields an effective heating rate, which is close to the empirically observed proton heating rate.} 
   {The observed scaling of fluctuation amplitudes is largely consistent with a picture of expanding Alfv\'{e}n waves that seed turbulence leading to dissipation. The expansion of the waves leads to the growth of wave-amplitudes, resulting in the formation of switchbacks.}

   \maketitle
%

\section{Introduction}
Fluctuations in the solar wind at scales much larger than ion-kinetic scales are often approximated in the framework of magnetohydrodynamics (MHD), which provides a complete description of the spatiotemporal evolution of the magnetic field, plasma flow velocity, and plasma density.  Linearization of the MHD equations in terms of small-amplitude plane waves identifies eigenmodes corresponding both propagating waves and non-propagating structures that share characteristics with observed solar wind fluctuations. The \cite{Alfven1942} mode, which is associated with proportional magnetic field and velocity fluctuations $\delta \mathbf{v}\propto\delta \mathbf{B}$, is known to agree well with observed polarization signatures of the solar wind \citep{Belcher1971}.

 While small amplitude approximations to MHD are useful in describing observed turbulence, the solar wind is often subject to large-amplitude fluctuations in the magnetic field with ($|\delta \mathbf{b}|/ |\mathbf{b}_0|\sim1$)  that maintain constant magnitude  $$|\mathbf{b}|=|\mathbf{b}_0+\delta \mathbf{b}|=const,$$ and appear spherically-polarized \citep{Goldstein1974,Lichtenstein1980,Riley1996,Gosling2009}. Constant magnitude fluctuations are highly prevalent in the inner-heliosphere observed by PSP \citep{DudokdeWit2020,Dunn2023}. These fluctuations share characteristics of finite- and large-amplitude Alfv\'{e}n waves, which maintain constant magnitude of the total magnetic field \citep{BarnesHollweg1974,Goldstein1974}, and thus appear spherically or arc polarized.  Specifically, large-amplitude fluctuations in the solar wind have a component parallel the mean field that maintains Alfv\'{e}nic correlations \citep{Matteini2015}; consequently, these fluctuations appear as ``one-sided'' enhancements in the solar wind velocity \citep{Gosling2009,Matteini2015}. The constraints imposed by the large-amplitude nature contrast the small-amplitude Alfv\'{e}n mode, which is a fluctuation purely perpendicular the mean magnetic field.  The large-amplitude, arc and spherically polarized Alfv\'{e}n modes are, like the small-amplitude (linear, plane-polarized) Alfv\'{e}n mode, exact solutions of MHD, and thus their ability to dissipate energy at large scales is negligible. Accordingly, even in these large amplitude states, which have recently been shown to occur in hybrid-kinetic models \citep{Matteini2024}, a turbulent cascade, or other nonlinearities \citep{Tenerani2020}, must drive energy in the large amplitude fluctuations to smaller, dissipative scales.

Recent observations from NASA's Parker Solar Probe mission (PSP) highlight large-amplitude inversions of the magnetic field, \citep{Bale2019,Kasper2019,DudokdeWit2020} many of which are strongly Alfv\'{e}nic \citep{Horbury2020,Laker2021,Mozer2020,Larosa2021}. These fluctuations, which have been termed ``switchbacks'' (SBs), are a signature of solar wind sources, and may be important in the energy budget of the solar wind heating as it expands into the heliosphere. The origin of large-amplitude SB fluctuations are a subject of significant debate and a comprehensive discussion is found in \cite{Raouafi2023}. In this Letter, we focus on the connection between spherically polarized waves and the in situ development of SBs.

Several authors have suggested that SBs likely arise in situ growth of Alfv\'{e}n waves in an expanding solar wind \citep{Shoda2021,Squire2020,Mallet2021,Johnston2022}, which explicitly couples the SB fluctuations to the large-amplitude Alfv\'{e}nic state. Here, we perform an observational analysis of the radial evolution of the spherically polarized, constant magnitude state. We identify a stream of fast wind in which the magnetic field fluctuations maintain strong spherical polarization as they propagate outwards; a stationary frame of the observed fluctuations is identified using de Hoffmann-Teller (dHT) analysis and found to well approximate the Alfv\'{e}n speed, suggesting that these spherically polarized waves are indeed Alfv\'{e}n waves. \cite{Bowen2025} have performed an analysis of the nonlinear turbulent interactions that occur in these states. Application of conservation of wave-action \citep{HeinemannOlbert1980,ChandranHollweg2009} demonstrates that the evolution of fluctuations in the stream is consistent with WKB expansion of large amplitude waves undergoing some amount of turbulent dissipation, which we find is consistent with proton heating rates. Expansion leads to the growth of normalized fluctuation amplitude $dB/B$, where $dB$ is understood as the rms amplitude of $\delta B$, leading to larger rotations in the field that resemble SBs. The growth of $dB/B$ leads to larger fractions of the solar wind magnetic field ``switching back'' at larger distances. These results suggest that SBs form in situ in the solar wind and that the decay of the large-amplitude state is a dominant contributor to solar wind heating.

\begin{figure}
    \centering
    \includegraphics[width=\columnwidth]{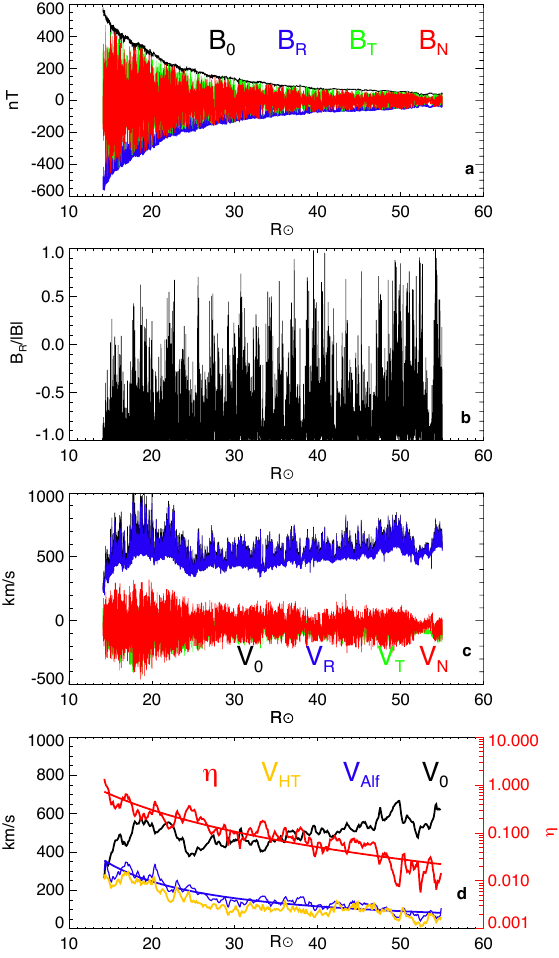}
    \caption{a) FIELDS magnetic field data in RTN coordinates, magnitude $|B|$ is shown in black. b) Radial component of magnetic field normalized to field magnitude $B_R/|B|$, large scale inversions in $B_R$ known as switchbacks are observed. c) SWEAP SPAN velocity measurements in RTN coordinates, $|V|$ is shown in black. d) Computed de Hoffmann-Teller frame speed ${V_{dHT}}$, Alfv\'{e}n speed $V_A$, and average wind speed $V_{sw}$ and $\eta=(V_A/V_{sw})^2$.}
    \label{fig:1}
\end{figure}
\section{Parker Solar Probe Observations}

We study a five-day stream from Parker Solar Probe's 10th perihelion encounter Nov, 16-21, 2021. During this interval, PSP traversed from a heliocentric distance of $54.9 R_\odot$ down to 14.3 $R_\odot$. The Encounter 10 stream is thought to be connected to a single coronal hole \citep{Badman2023} and the dynamical outer-scale evolution has been previously studied by \cite{Davis2023} and \cite{Huang2023}, who report the growth of turbulence from an outer-scale population of Alfv\'{e}nic fluctuations. Magnetic field data is obtained from the electromagnetic PSP/FIELDS experiment \cite{Bale2016} and proton moments from the PSP/SWEAP investigation's \citep{Kasper2016} electrostatic analyzer \citep{Livi2022}.

We separate the interval into $N=355$ one hour sub-intervals. Each sub-interval overlaps its neighbors by 2/3, e.g. each sub interval starts 20 minutes after the start of the previous sub interval. Overlapping was used to increase statistics. In each subinterval we compute the mean magnetic field vector $\mathbf{B}_0=\langle \mathbf{B} \rangle$, its magniude $B_0=|\mathbf{B}_0|$. We also compute mean magnetic field magnitude $B=\langle|{\mathbf{B}}|\rangle,$ the mean solar wind speed $V_{sw}=|\langle {\mathbf{V}}_{p}\rangle|$, and the Alfv\'{e}n speed $V_A=B_0/\sqrt{\mu_0\rho_0}$, where $\mu_0$ is the magnetic permeability and $\rho_0$ is the mean mass density of each hour computed from quasithermal noise densities \citep{Moncuquet2020, Pulupa2017}. We define
     $\eta=(V_A/V_{sw})^2$
following \cite{ChandranHollweg2009}.

Figure 1(a-c) show the vector magnetic field, the normalized $B_R/|B|$ compnent, and solar wind velocity vector of the interval respectively as a function of radial distance $R$. Figure 1(d) shows estimates of $\eta$, $V_A$, and $V_{sw}$ computed in each sub-interval, along with the power-law fits to each quantity in the form $\bar{\eta}= \eta_0 (r/r_0)^{\alpha_\eta}$. 

Figure 1(d) additionally shows the de Hoffmann Teller (dHT) frame computed in each hour-long interval. The dHT frame minimizes the rms value of the convected electric field ${\bf{E}}=-{\bf{v}} \times {\bf{B}}$ \citep{1998ISSI} corresponding to aligned magnetic and velocity fluctuations. The frame with zero-electric field corresponds to a stationary frame for electromagnetic fluctuations with no parallel electric field, such as Alfv\'{e}n waves. We obtain the dHT through minimizing $E^2$ for each interval:
\begin{align}\label{eq:dht} 
E^2=\sum_i \left(({\bf{V}}_{dHT} -{\bf{v}}_i)\times {\bf{B}}_i\right)^2,\\
\frac{\partial{E^2}}{\partial{\bf{V}}_{dHT}}=0.
\end{align}
The dHT frame, ${\bf{V}}_{dHT}$, is invariant under multiplicative scalings of ${\bf{B}}$, and is thus not sensitive to Alfv\'{e}nic normalization. \cite{Matteini2015} previously have used the dHT frame to study the minimization of the convected electric field via spherical polarization. Additionally, the dHT has been used to study and classify switchbacks \citep{Horbury2020}. Recently, \cite{Agapitov2023} demonstrated that a shared dHT exists between the SB fluctuations and the surrounding solar wind. Figure \ref{fig:1}(d) shows that the magnitude of the dHT frame,  ${|\bf{V}}_{dHT}|$ follows the Alfv\'{e}n speed closely, which suggests that the outer, 1 hour long scales consist predominantly of Alfv\'{e}nic fluctuations.

In each interval we also compute the RMS fluctuation quantitiy of the magnetic field, $dB$; we furthermore compute rms quantities of the magnetic field perpendicular to the mean field, $dB_{\perp}$, and parallel the mean field, $dB_{\parallel}$. To highlight the spherically polarized nature of this stream,  we fit the vector magnetic field data in each sub-interval to a spherical shell of constant magnetic field magnitude $B^{sph}$, using linear-least square optimization techniques. We project each vector magnetic field measurement onto the spherical shell to produce a spherically polarized $\mathbf{B}^{sph}$, that points parallel to the measured $\mathbf{B}$ at each time, but with a constant magnitude $|\mathbf{B}^{sph}|$. We compute the RMS fluctuation quantities of the spherically polarized magnetic field perpendicular to the mean field, $dB^{sph}_{\perp}$, parallel the mean field $dB^{sph}_{\parallel}$. Furthermore, we compute the residuals $\mathbf{B}_C=\mathbf{B}-\mathbf{B}^{sph}$, corresponding to compressible fluctuations off of the spherical shell, and report their average values for each interval as ${dB_C}=\sqrt{\langle\mathbf{B}_C^2\rangle}$. As the spherically polarized $\mathbf{B}^{sph}$ vector is parallel to $\mathbf{B}^{sph} $ at each measurement, the deviation off the sphere is essentially equal to  $|{\mathbf{B}}|-\langle|{\mathbf{B}}|\rangle$, such that  $dB_C$  is equivalent to the rms of $d|B|$.

\begin{figure}
    \centering
    \includegraphics[width=\columnwidth]{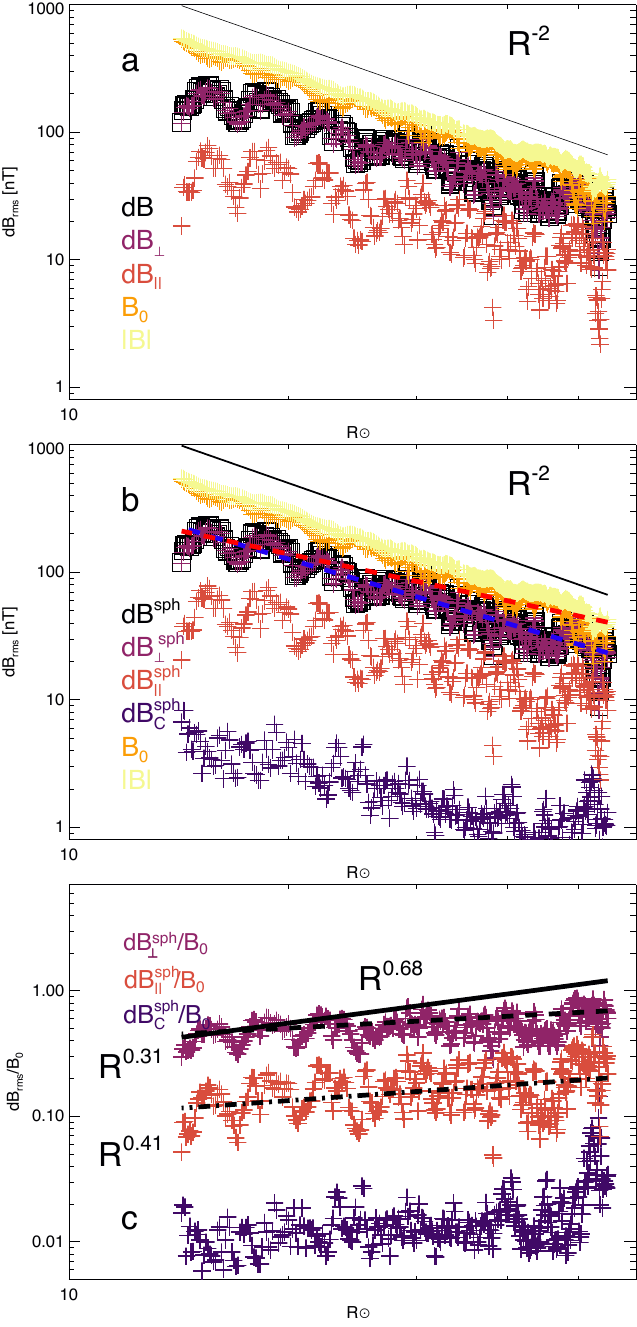}
    \caption{a) Root mean squared $dB$, $dB_\perp$ and $dB_\parallel$ (black, maroon, red); magnitude of mean magnetic field $B_0$ orange. b) Spherically confined fluctuations  $dB$, $dB_\perp$ and $dB_\parallel$ (black, maroon, red) where all fluctuations maintain constant magnitude $|B_{sph}|$. The rms residual fluctuations between the total field and spherically polarized field, $dB_C$ are shown in dark blue. Red dashed line shows wave amplitude from conservation of wave action with no dissipation, blue line shows best fit to $dB^{sph}$. c) Normalized amplitudes $dB_\perp^{sph}/B_0$, $dB_\parallel^{sph}/B_0$ and $dB^{sph}_{C}/B_0$. Best fit lines to   $dB_\perp^{sph}/B_0$ and $dB_\parallel^{sph}/B_0$ are shown as well as the scaling expected for conservation of wave action.}
    
    \label{fig:2}
\end{figure}

Figure 2(a) shows the radial dependence of RMS quantities $dB_{\perp}$, $dB_{\parallel}$, while Figure 2(b) shows the spherically confined quantities $dB^{sph}_{\perp}$, $dB^{sph}_{\parallel}$ alongside ${dB_C}$ and ${B_0}$. 
The spherically constrained $dB^{sph}_{\perp}$, $dB^{sph}_{\parallel}$ are nearly identical to the unconstrained quantities indicating the strong spherical polarization throughout the stream. The deviation off of the sphere is ${dB_C}$, and significantly smaller than the perpendicular or parallel rms fluctuations, indicating that the data in each interval is well fit by a sphere.

 Given the strong spherical polarization, which includes significant fluctuations parallel the mean and that maintains a stationary dHT frame that is consistent with the Alfv\'{e}n speed,  we argue that the spherically polarized magnetic field fluctuations correspond to an outward propagating, finite-amplitiude, Elsasser mode $\mathbf{z}^{\pm}=\mathbf{v}\pm\mathbf{b}$ with $\mathbf{z}^+ \approx 2\mathbf{b}^{sph}$, where $\mathbf{b}^{sph}= \mathbf{B}^{sph}/{\sqrt{\mu_0\rho}}$. We use a linear least square fit in log-log space of $dB^{sph}$  and radius $R/R_\odot$ to approximate power-law scalings $\bar{dB}^{sph}= dB_0^{sph} (R/R_\odot)^{\alpha_{dB}}$; we find $\alpha_{dB}=1.67$, which is shown in blue in Figure 2(b).

The quantity $\eta$ is used to define the quantities $\mathbf{g^{\pm}}$ from the Elsasser fluctuations $\mathbf{z^\pm}$ as 
\begin{equation}
    \mathbf{z}^{\pm}= \mathbf{g}^{\pm}\frac{\eta^{1/4}}{1\pm\eta^{1/2}}. \label{eq:gz_eta}
\end{equation}
The quantity ${g^+}^2$, which corresponds to the outward-propagating Elsasser variable $\mathbf{z}^+$, is ideally conserved in the absence of dissipation and heating, and is conserved even for large-amplitude fluctuations, so long as they remain spherically polarized \cite{Hollweg1974}. We compute ${dz}^{+}= 2{d{B}}^{sph}/\mu_0{\rho}$ from 
a linear least square fit in log space of the form $\bar{dz}^{+}= dz^+_0(R/R_\odot)^{\alpha_{dz^+}}$.
Using $dz^{+}_0$, 
we define $dg^{+}_0$. 
We compute dissipationless  scalings for $\bar{g}'(R)$ and $\bar{z}'(R)$ from Equation \eqref{eq:gz_eta}, ${g^+_0}$, and the scaling of $\bar{\eta}$; 
the dissipationless $dB'(R)$, with approximate power law scaling $\alpha_{dB'}=1.22$, is similarly computed and plotted in red in Figure 2(b).

Figure 2(c) shows normalized amplitudes $dB/B_0$ that grow with radius. We compute a power law fit to $dB_\perp^{sph}/B_0$ with a scaling of $R^{0.31}$, the dissipationless scaling of ${dB^{sph}}'/B_0$, which is how the amplitudes would grow without heating, is found to be $R^{0.68}$. We compute a power law fit to $dB_\parallel^{sph}/B_0$ with a scaling of $R^{0.41}$. Compressible fluctuations, $dB_C/B_0$, remain at $\approx0.01$ of the mean magnetic field up until about $50R_\odot$, when there is significant growth. 

The observed deviation of $dB$ from the dissipationless $dB'(R)$ indicates the decay of large amplitude fluctuations, likely associated with dissipation and extended heating of the solar wind. Following \cite{ChandranHollweg2009} and \cite{ChandranPerez2018}, 
we obtain a decay rate for the large scale Alfvénic fluctuations as:

\begin{align}\label{eq:qwv}
   Q(R) = - \frac{\rho_0}{4(1+\eta^{1/2})}{g^+}^2\frac{d\bar{V}_A}{dR}
\end{align}
where the derivative $\frac{d\bar{V}_A}{dR}$ is obtained from the power-law fit $\bar{V}_A$. Importantly, this estimate of $Q(R)$ in Eq.~\ref{eq:qwv} should correspond to the total turbulent heating rate that is deposited in both ions and electrons.

Figure 3(a) shows the estimated heating rate from the conservation of wave-action $Q(R)$. We compare this quantity to the proton and electron heating rates $Q_{e,p}$, as well as the turbulent cascade rate $\epsilon$. Recent work has highlighted that analysis of the proton heating rate is best obtained via direct analysis of particle properties \citep{Zaslavsky2023,Mozer2023}. For each individual sample in the stream we compute the pressure perpendicular and parallel the magnetic field $p_\perp$ and $p_\parallel$ and the constants $C_\parallel= p_\parallel B^2/n^3$ and $C_\perp =p_\perp/n B$,
which are related to the heating terms $Q_{p\perp}$ and $Q_{p\parallel}$ via 

\begin{align}
    p_\parallel\frac{d\text{ln}{C_\parallel}}{dt}=Q_{p\parallel}\\
p_\perp\frac{d\text{ln}{C_\perp}}{dt}=Q_{p\perp}
\end{align}
\citep{Chew1956,Zaslavsky2023,Mozer2023}. The anisotropic pressures are computed from the PSP SPAN-I Temperature tensor \citep{Livi2022}, which is rotated into field aligned coordinates. We include only SPAN-I observations where more than 60 energy bins had finite, non-zero counts, which is intended to remove observations where the plasma is out of SPAN's limited field of view. We additionally implemented drifting bi-Maxwellian fits of a core and beam population to the SPAN-i data \citep{Bowen2024b} and computed effective parallel and perpendicular pressures following \cite{Klein2021}; no significant deviations were found when implementing the moments versus the fit approximations to the SPAN-i data.

We compute power-law fits $\bar{C_\perp}(R)$ and $\bar{C_\parallel}(R)$. From which $Q_p$ is estimated as \begin{align} \label{eq:QCGL} Q_p={p}\frac{d\text{ln}{\bar{C}}}{dt},\end{align} and the time derivative is approximated using the convective derivative $d/dt= V_{sw}d/dR$.  Figure 3(a) shows $Q_p=Q_{p\perp} +Q_{p\parallel}$ to demonstrate general agreement between the total proton heating observed from the plasma parameters versus the expected heating from the deviations from conservation of wave action in Equation \ref{eq:qwv}. There is some subtlety to the use of $|B|$ or $B_0$ in the CGL approach \citep{Marsch1983,Perrone2019,Zaslavsky2023}; though we leave in-depth discussion to CGL invariants to future discussion, we do note that the use of $|B|$ or $B_0$ in Equation \ref{eq:QCGL} results in only a 10\% difference in heating rates.

We additionally consider the electron heating through scaling of the QTN estimate of the electron temperature, \cite{Moncuquet2020}, which is available up to approximately $40R_\odot$. Though we do not have measurements of electron temperature anisotropy to perform the CGL analysis, we implement the following electron heating formula \begin{align} \label{eq:QE_cran}
    Q_e= \frac{3}{2}n_eV_{sw}k_B\frac{dT_e}{dR} - V_{sw}k_BT_e\frac{dn_e}{dR}
\end{align} \citep{Cranmer2009,Bandyopadhyay2023}, where terms corresponding to collisonal heating and the Parker spiral effects are omitted due to their minimal effects. We have also omitted consideration of heat flux which is likely negigible compared to contributions from wave heating \cite{Halekas2023} in these faster streams. Though heat-flux is likely important for the electron energy budget in the inner helioshere \citep{Halekas2020,Bercic2020}. In evaluating Equation \ref{eq:QE_cran}, we use direct measurements of $n_e$ and $T_e$ and derivatives of the power law fits to estimate gradients ($d\tilde{T_e}/dR$ and $d\tilde{n_e}/dR$) as directly computing gradients results in significant noise lacking in clear interpretation.
A corresponding equation for $Q_p$ similar to Eq.~\ref{eq:QE_cran} exists, though the results are indistinguishable from the analysis obtained via consideration of CGL invariants.  Figure 3(a) shows that $Q_e$ is substantially lower than $Q_p$, the ratio of $Q_p/Q_e$ is shown in Figure (3b) and a PDF is shown in Figure 3(c), the median value of $Q_p/Q_e$ is approximately 4, such that only $\approx 20\%$ of energy is dissipated into electrons. 

These quantities can further be compared to the turbulent cascade rate $\epsilon$, which we compute using the \cite{PolitanoPouquet1998} 3rd order scaling relations:
\begin{align}
   Y^\pm{(l}) &=-\frac{4}{3}\epsilon^{\pm}l\\
      Y^\pm{(l}) &=\langle\hat{l}\cdot\Delta\mathbf{z}^\mp|\Delta\mathbf{z}^{\pm}|^2\rangle
\end{align}
where $\Delta \mathbf{z}^\pm$ refer to two point increments in the Elsasser variables and the Taylor hypothesis is used to convert from time-lags $\tau$ to the spatial scale $l$ as $\tau \langle {\mathbf{V}}_{p}\rangle=l\hat{l}$. The total energy cascade rate is equal to $\epsilon=(\epsilon^+ +\epsilon^-)/2$. For each interval, we compute $\epsilon^\pm$ for a range of lags between 1 and 180 seconds, which is in inertial range at all distances, and take the total $\epsilon$ as the average over these scales. Figure 3(a) shows that $\epsilon$ follows the heating rate $Q_p$ and WKB dissipation rate $Q(R)$ relatively closely. 

Figure 3(b) shows the ratio of $\epsilon$ to the WKB dissipation rate $Q$ as well as the ratio $Q_p/Q$ and the ratio $Q_\perp/Q_\parallel$. Probability distribution functions (PDFs) of log-base-10 of these quantities are shown in Figure 3(c). These quantities are notoriously hard to measure with great precision and accuracy, to understand the errors on these measurements we compute variances of the PDFs in $\text{Var[}\text{Log}_{10}Q_p/Q] =0.34$ ,$\text{Var}[\text{Log}_{10}Q/\epsilon]=0.41$, $\text{Var}[\text{Log}_{10}Q_p/\epsilon]=0.45$. 
Assuming that the error in each of these quantities is approximately equal and normally distributed, (${\sigma_Q} \approx {\sigma_\epsilon}{\approx\sigma_{Q_p}})$, standard propagation of uncertainties can be used to estimate the measurement error from the variances of the distributions under the assumption that $Q=Q_p=\epsilon$ holds true--i.e. dispersion in the data shown in Fig 3 is due solely to error. Following quadrature error analysis, the fractional error in each measured quantitiy is defined as $\sqrt{2}\sigma_x/x =\sigma_{PDF}/\text{ln{10}}$, where $\sigma_{PDF}$ is the uncertainty measured from the PDF in Figure 3(c). The measured variances imply a fractional error $\sigma_Q/Q \approx\sigma_\epsilon/\epsilon$ that ranges between 0.9-1.2. Essentially, this error analysis indicates that the dispersion in the measured ratios is consistent with perfectly equal cascade rates, wave-dissipation, and proton heating rates, if the fractional errors on the measurement are on order unity. While these estimation techniques contain significant error, the measurements and scaling provide good evidence that the decay of large scale waves matches the turbulent cascade and net heating rates.

Combined with the observations of spherical polarization, strong Alfvénic nature of the outer scale fluctuations, the in Figures 2 \& 3 are consistent with the interpretation of large-scale Alfv\'{e}n waves that grow in normalized amplitude $dB/B_0$ due to expansion effects, but also decay through a turbulent cascade, and dissipate their energy into the solar wind plasma consistent with recent results \citep{Rivera2024}.

\begin{figure}
    \centering
    \includegraphics[width=\columnwidth]{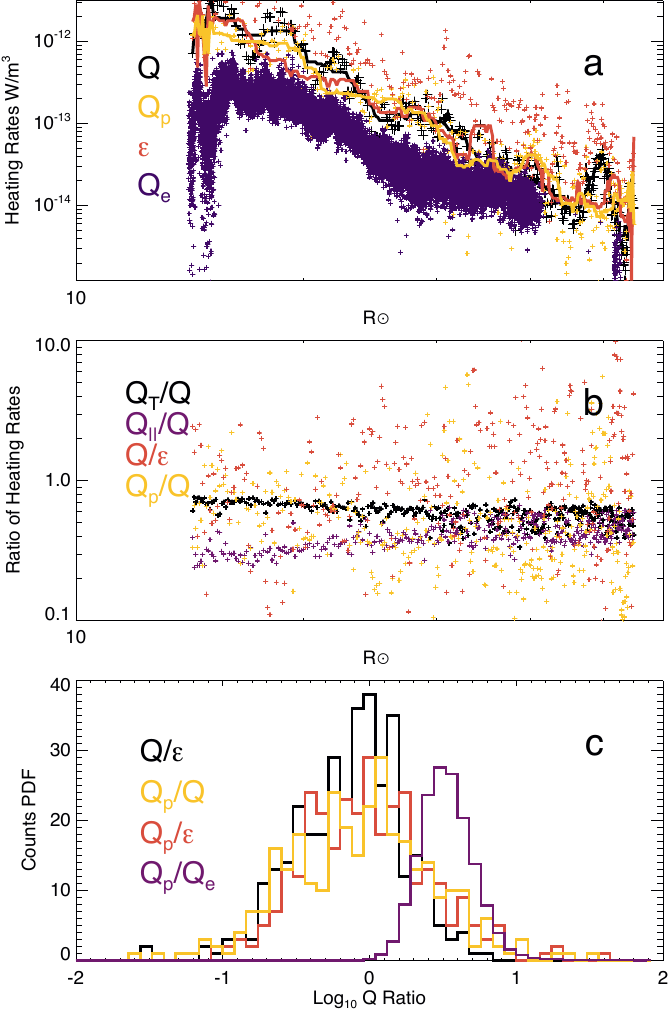}
    \caption{a) Radial scaling of dissipation rates computed from WKB-expansion of outer scale fluctuations $Q$ (black) from \cite{Chandran2009}, CGL estimate of proton heating $Q_p$ (orange) followng \cite{Zaslavsky2023}, 3rd-order cascade rate $\epsilon$ \citep{PolitanoPouquet1998}, and electron heating $Q_e$ \citep{Cranmer2009}; sold lines show moving-window median value. b) Ratios of dissipation quantities, $Q_{p\perp}/Q$ (black) and $Q_{p\parallel}/Q$ (purple), $Q/\epsilon$ (red), $Q_p/Q$ (orange). c) PDFs of ratios $Q/\epsilon$ (black),$Q_p/Q$ (orange), $Q_p/\epsilon$ (red), $Q_p/Q_e$ (purple).}
    \label{fig:enter-label}
\end{figure}

\section{Growth of Switchbacks}
\begin{figure*}
    \centering
    \includegraphics{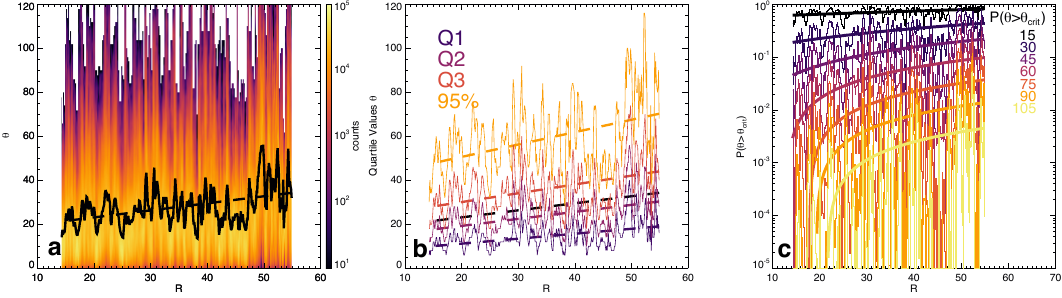}
    \caption{a) Distribtion of $\theta$, the angle between the magnetic field and the mean field diretion as function of solar radius. Black line shows the mean $\theta$ at each radius, with a linear least square fit line shown in black. b) Quartiles and 95th percentile of distribtuion shown in panel (a). Least square fit lines are shown for each percentile. c) Probability for $\theta>\theta_{crit}$ for a range of $\theta_{crit}$ between 15 and 105 degrees; larger $\theta_{crit}$ is shown in lighter colors; linear least square fit trend lines are plotted for each $\theta_{crit}$.}
    \label{fig:3}
\end{figure*}
For a spherically polarized magnetic field, with magnitude $B$ fluctuations in the field $\delta B$ correspond simply to rotations around the sphere with $\delta B^2= 2B^2 - 2B^2 cos\psi$, where $\psi$ is the rotation angle of the field. 
Generalized increased growth in normalized rms amplitude $dB/B$, which Fig.~2 demonstrates occurs with radial distance, naturally leads to an increase in the size of angle $\psi$. It has been argued that large-angle rotations that invert the magnetic field in the solar wind, forming a SB, in Alfv\'{e}nic intervals can be formed simply through the growth of the fluctuation amplitude via expansion \citep{Matteini2015,Squire2020,Mallet2021}. It is clear that together, the growth of $dB/B_0$, along with the constrained spherical polarization demand an increase in large rotation angles $\psi$, which can invert the field.


For each hour interval we compute the average field and rotate the data into field aligned coordinates (FAC) , defined as $(x,y,z)=(B_{\perp1},B_{\perp2},B_{\parallel})$. We define the angle $\theta$ as the angle between the magnetic field at each time and the average field direction, which is defined as the time average over the hour. Figure 4(a) shows the distributions of measured $\theta$ for each interval as a function of $R$. A value of $\theta=0$ corresponds to the magnetic field pointing along the mean field direction, while an angle of $\theta=180$ is a complete inversion, a SB or inversion of the field relative to its mean direction corresponds to angles of $\theta=90^\circ$. We plot the average $\theta$ as well as the result of linear least square regression to average $\theta$. The mean value of $\theta$ increases with distance, indicating that, on average, the magnetic fields deviation from the mean field direction grows with distance.

In Figure 4(b) we plot quartiles of the distribution of $\theta$ from each interval as a function of $R$ as well as the 95th percentile of the distribution of $\theta$. We show a best-fit line to each quartile and the 95th percentile: the distributions systematically move towards higher average values of $\theta$, which is consistent with the growth of the fluctuations from expansion.  We also plot the best-fit line to the average $\theta.$ Finally, Fig 4(c) further highlights the growth of switchbacks by showing, for each interval, the probability that the angle $\theta$ is larger than a critical angle $\theta_{crit}$, which is varied from $\theta_{crit}=15^\circ$ to $\theta_{crit}=105^\circ$. 
For each $\theta_{crit}$ we fit linear best-fit lines. The slopes of the best fit lines clearly show increasing trends, indicating that the probability for $\theta>\theta_{crit}$ increases with distance. For inversion angles of $\theta=90\circ$, there is a very low probability $P(\theta<90^\circ) <0.1\%$ at $20R_\odot$, whereas $P(\theta<60^\circ) \approx 1\%$ at $20R_\odot$. At $55R_\odot$ $P(\theta<90^\circ)\approx 1\%$  and $P(\theta<60^\circ)\approx 10\%.$ The structure of the least square trend lines, indicating switchback growth, are similar to numerical simulations of switchback formation via expansion \citep{Johnston2022}.

\section{Discussion}
Magnetic field inversions have been observed in the heliosphere for many years \citep{McCracken1966,Balogh1999,Gosling2009}. However, recent observations from PSP have brought significant focus on these structures and their use in constraining solar wind sources \citep{Bale2023,Raouafi2023b}, and their role in heating the solar corona and wind \citep{Woolley2020,Laker2021,Tenerani2020}. Understanding the formation of these structures is fundamental to successfully employing them as a constraint on solar wind sources and understanding their interplay with heating. In this Letter, we present evidence that many SBs may be formed in-situ via the effects on expansion on large-scale outward propagating Alfv\'{e}n waves \citep{Squire2020,Shoda2021,Mallet2021,Johnston2022,Matteini2024}.

Our results suggest that the outer-scale fluctuations are consistent with a population of large-amplitude spherically polarized fluctuations. The stationary frame of the outer-scale fluctuations, measured through the de Hoffmann-Teller frame, is consistent with the Alfv\'{e}n speed, suggesting that the large amplitude, constant $B$ structures are likely Alfv\'{e}n waves propagating away from the sun \citep{Gosling2009,Matteini2015}, and are often found in the near-sun solar wind \citep{Dunn2023}.

Application of conservation of wave action \citep{HeinemannOlbert1980,ChandranHollweg2009} demonstrates that the empirically observed wave amplitudes grow as the solar wind expands. As the waves grow through expansion, the normalized amplitudes increase, leading to greater fractions of the observed fluctuations having amplitudes capable of "switching back" the magnetic field. Importantly, the growth of the amplitudes is not entirely consistent with conservation of wave-action indicating that some energy from the waves must be lost \citep{ChandranPerez2018}. However, empirical measurement of proton heating rates via deviations from adiabatic expectations \citep{Chew1956,Zaslavsky2023,Mozer2023} reveals levels of proton heating roughly consistent with heating rates obtained from both consideration of conservation wave action at outer-scales as well as the turbulent energy flux \citep{PolitanoPouquet1998}. While there are significant uncertainties on calculations of the turbulent cascade rate and heating rates, the general correspondence and similar scalings observed between the dissipated energy, turbulent cascade rate, and proton heating rates suggest that these processes are closely intertwined. These observations are important in understanding the expanding solar wind and suggest that the onset of the turbulent cascade and solar wind heating are important in regulating the growth of switchbacks. Increased amounts of heating would result in lower growth rates of $dB/B$, which may inhibit the growth of switchbacks through expansion. The kinetics of proton heating in this stream were recently studied by \cite{Bowen2024b}, where significant cyclotron resonant resonant was found but partition in ion-electron heating rates was not considered. The observation of $Q_p/Q_e$ >1 we report here is consistent with cyclotron resonant heating mechanisms as a means for dissipating turbulence previously shown in \cite{Bowen2024a}. While we have omitted the electron heat flux, these measurements, which indicate residual electron heating in excess of adiabatic evolution, are consistent with the overall magnitude of heating found by \cite{Stevrak2015}; however, we have not considered the contribution to the heat flux, which \cite{Stevrak2015} found to be larger in magnitude to the $Q_e$ that we consider, and furthermore is negative in sign and thus significantly affect the electron heating. Further analysis of electron heating should use SPAN-e data \cite{Halekas2020} to fully analyze the radial electron distributions to understand the subtleties in the partition of ion and electron heating. Similarly, a number of subtleties exist in the application of CGL \citep{Zaslavsky2023}, future studies on radial scans of PSP, which can provide analysis of solar wind evolution in a single stream are important to understanding the connections between specific microphysical heating processes with non-adiabatic evolution of the observed VDFs. 

Further work is necessary to consider the in situ development of the SBs through other mechanisms: for instance, the Kelvin Helmholtz Instability (KHI) a leading alternative to the in situ development via expansion \citep{Mozer2020,Ruffolo2020}. \cite{Agapitov2023} have recently studied KHI in the presence of SBs and note that observed switchbacks seem stable to KHI, though that they may be regulated by the instability. Additionally, a significant body of work has discussed the generation of switchbacks through reconnection much closer to the solar surface \cite{Drake2021,Bale2021,Fargette2021,Bale2023}. While our results do not preclude the generation of these features close to the sun, our analysis clearly supports a role of expansion and in situ formation of switchbacks. Patches of switchbacks \citep{Bale2021,Fargette2021,Shi2022} have gained particular interest in the community as the modulated patch-size seems to correspond to super-granulation scales. If this modulation of the switchbacks does indeed correspond to features on the solar surface, it remains an important question to understand how the patch-like modulation evolves under expansion and dissipation.

Many open questions remain regarding switchbacks: their origins, relation to heating, and fundamental relations and impact on solar wind turbulence \citep{Raouafi2023}. In this brief letter, we have attempted to clearly demonstrate evidence for the natural growth of SBs through expansion. Indeed, the spherically polarized Alfv\'{e}nic state is an equilibrium solution to the MHD equations and it is certainly possible that many physical processes may inevitably relax to this state.

\begin{acknowledgements}
TAB acknowledges support from 80NSSC21K1771. BDGC acknowledges the support of NASA grant 80NSSC24K0171. This work was supported by the ISSI "Magnetic Switchbacks in the Young Solar Wind" workshop.
\end{acknowledgements}

%
%
\bibliographystyle{aa} 
\providecommand{\noopsort}[1]{}\providecommand{\singleletter}[1]{#1}%

\end{document}